\documentclass[aps,10pt,twocolumn,prl,floatfix]{revtex4-1}
\usepackage{graphicx}
\usepackage{amssymb,amsmath,tabularx}
\usepackage{bm}
\usepackage[caption=false]{subfig}

\begin{document}

\title{Coevolution Maintains Diversity in the Stochastic ``Kill the Winner" Model}

\author{Chi Xue}
\affiliation{Department of Physics and Center for the Physics of Living Cells, University of Illinois at Urbana-Champaign \\
Loomis Laboratory of Physics, 1110 West Green Street, Urbana, Illinois 61801-3080, USA}
\affiliation{Carl R. Woese Institute for Genomic Biology and Institute for Universal Biology, University of Illinois at Urbana-Champaign \\
1206 West Gregory Drive, Urbana, Illinois 61801, USA }

\author{ Nigel Goldenfeld}
\email{nigel@uiuc.edu}
\affiliation{Department of Physics and Center for the Physics of Living Cells, University of Illinois at Urbana-Champaign \\
    Loomis Laboratory of Physics, 1110 West Green Street, Urbana, Illinois 61801-3080, USA}
\affiliation{Carl R. Woese Institute for Genomic Biology and Institute for Universal Biology, University of Illinois at Urbana-Champaign \\
    1206 West Gregory Drive, Urbana, Illinois 61801, USA }

\date{\today}

\begin{abstract}
The ``Kill the Winner" hypothesis is an attempt to address the problem
of diversity in biology. It argues that host-specific predators control
the population of each prey, preventing a winner from emerging and thus
maintaining the coexistence of all species in the system. We develop a
stochastic model for the ``Kill the Winner" paradigm and show that the
stable coexistence state of the deterministic ``Kill the Winner" model
is destroyed by demographic stochasticity, through a cascade of
extinction events. We formulate an individual-level stochastic model in which
predator-prey coevolution promotes high diversity of the ecosystem by
generating a persistent population flux of species.
\end{abstract}


\maketitle

The high diversity of coexisting species in most ecosystems has been a
major puzzle for more than 50 years.  In a classic paper, Hutchingson
articulated the so-called Paradox of the Plankton for the case of
marine ecosystems \cite{hutchinson1961paradox}: why do many species of
plankton that feed on the same nutrients coexist, instead of one
species outcompeting all the others?  This latter expectation might
seem to be intuitive, and has also been formulated precisely as the
so-called competitive exclusion principle
\cite{hardin1960competitive_exclusion}. The Paradox of the Plankton is
not limited to marine ecosystems, but has been generalized to
terrestrial systems and expressed as the biodiversity paradox
\cite{clark2007biodiversity_paradox, wilson1990mechanisms}.

The various tentative resolutions of the paradox can be divided into
two classes \cite{wilson1990mechanisms, roy2007paradox_resolution}. In
the first, the ecosystem is argued to have failed to reach a fixed
point equilibrium state in which the competitive exclusion principle
applies, due to temporal or/and spatial factors. For example, the time
needed for the system to reach equilibrium might be much longer than
the time over which the system undergoes significant changes in its
boundary conditions, such as weather \cite{levins1979coexistence}.
Spatial heterogeneity can increase the global diversity of the system
by maintaining local patches that each obey the competitive exclusion
principle but globally support the coexistence of multiple species
\cite{richerson1970contemporaneous, levin1974dispersion}. In the second
class of resolutions, interactions such as predation, in conjunction
with competitive exclusion, promote the co-existence of species through
time-dependent or stochastic steady states \cite{scheffer2003plankton,
roughgarden1975species, thingstad1997KtW, vetsigian2011structure}. One widely celebrated
example of this behavior, which is seen in both natural ecosystems as
well as some laboratory systems such as chemostats
\cite{fernandez1999stable,doebli2017}, is the continual succession of
different community members known as ``Kill the Winner" (KtW)
dynamics \cite{thingstad1997KtW, thingstad2000elements,
winter2010KtW_revisit}.

In the KtW hypothesis \cite{thingstad1997KtW, thingstad2000elements,
winter2010KtW_revisit} there are two groups of resource consumers, for
example bacteria and plankton. The plankton community generally has a
lower efficiency of resource usage than bacteria. They remain in the
system, only because a protozoan consumes the bacteria non-selectively
and thus limits the bacterial population, leaving room for plankton to
thrive. Inside the bacterial community, different strains have distinct
growth rates. They coexist, with no dominating winners, due to
host-specific viruses controlling the corresponding strains.
This results in two layers of coexistence through KtW dynamics
(bacteria-plankton coexistence and bacterial strain coexistence),
rested like Russian dolls \cite{winter2010KtW_revisit}.

The original KtW model \cite{thingstad1997KtW, thingstad2000elements}
was formulated as deterministic Lotka-Volterra type equations for the
species biomass concentrations. The high diversity of the system is
exhibited in the steady state where multiple species coexist with
positive biomass; these calculations assume that the system is
spatially homogeneous and that the number of individuals is large
enough that it is valid to use a continuous density to describe the
population. However, this is not appropriate when the population is
finite, because large fluctuations are able to drive the system towards
extinction, an outcome that cannot be captured by a continuous density
that is allowed to become arbitrarily small.  Requiring the population
size to be integer-valued leads inexorably to shot noise, referred to
in the ecological context as demographic stochasticity.

The purpose of this Letter is to explore the effect of demographic
stochasticity on the ``Kill the Winner" paradigm and demonstrate that
the stochasticity causes the coexistence steady state in the
deterministic KtW model to break down through a cascade of extinctions,
leading to a loss of diversity.  This extinction can be avoided by
allowing the predators and prey to coevolve.  We propose a stochastic
model of this coevolution and show that it generically maintains the
diversity of the ecosystem, even in the absence of spatial extension.
Our results strongly suggest that diversity reflects the dynamical
interplay between ecological and evolutionary processes, and is driven
by how far the system is from an equilibrium ecological state (as could
be quantified by deviations from detailed balance). The surprisingly
deep role of demographic stochasticity uncovered here is consistent
with earlier demonstrations of strong fluctuations and qualitatively
new phenomena in ecosystems where local populations are small. For
example, it is now well-understood how individual-level minimal models
can account for a wide variety of ecological phenomena, including
large-amplitude persistent population cycles
\cite{mckane2005predator-prey}, anomalous phase shifts due to the
emergence of mutant sub-populations
\cite{yoshida2003rapid,shih2014rapid_evolution}, spatial patterns
\cite{tauber2012population, butler2009robust, biancalani2010stochastic,
biancalani2017non_normality} and even reversals of the direction of
selection \cite{constable2016demographic} without requiring overly
detailed modeling of inter-species interactions.

\smallskip
\noindent\textit{Model:-} The key component of the KtW hypothesis is
that, for each resource competitor, there is a corresponding predator
that can prevent it from becoming a dominant winner. The Russian
doll-like hierarchy is not essentially important for the basic idea. Thus we
will focus on only a single layer of KtW interaction,
the host-specific viral infection, and ignore the multilevel structure.

We write down the individual reactions for a simplified system of $m$
pairs of prey and predators, which we will take to be bacteria and
viruses (phages), as follows:
\begin{subequations} \label{multiPP_reactions}
\begin{align}
    X_i &\rightarrow  2 X_i,  &\text{with rate }b_i, \\
    X_i + X_j &\rightarrow  X_i,  &\text{with rate }e_{ij}, \\
    Y_i + X_i &\rightarrow  ( \beta_i + 1 ) Y_i,  &\text{with rate }p_i, \\
    Y_i &\rightarrow \varnothing,  &\text{with rate }d_i.
\end{align}
\end{subequations}
Here $i, j = 1, 2, \dots, m$ are strain indices. Bacterial individuals
$X_i$, have strain-specific growth rate $b_i$. They compete with each
other for an implicit resource with strength $e_{ij}$. Viruses of the
$i$th strain $Y_i$, infect the corresponding host $X_i$ with rate $p_i$
and burst size $\beta_i$, and decay to nothing $\varnothing$ with rate
$d_i$. These reactions form a minimal model, which we refer to as the
generalized KtW model, and ignore many biological details that are
present in ecosystems.

The corresponding mean-field rate equations are shown below.
\begin{subequations} \label{multiPP_ODE}
\begin{align}
    \dot{B_i} &= b_i B_i - \sum_{j=1}^m e_{ij} B_i B_j - p_i B_i V_i, \\
    \dot{V_i} &= \beta_i p_i B_i V_i - d_i V_i.
\end{align}
\end{subequations}
The dot operator stands for the time derivative. $B_i$ and $V_i$
represent the density of the $i$th bacterial and viral strains
respectively. For simplicity, we set $e_{ij} \equiv e$.

The nonzero steady state of Eq. (\ref{multiPP_ODE}) is given by the following equations.
\begin{subequations} \label{multiPP_steady_state}
\begin{align}
    B_i^* &= \frac{d_i}{\beta_i p_i}, \\
    V_i^* &= \frac{1}{p_i} \left( b_i - e\sum_{j=1}^m B_j^* \right).
\end{align}
\end{subequations}
Linear stability analysis shows that the steady state Eq.
(\ref{multiPP_steady_state}) is exponentially stable, with all eigenvalues
of the linear stability matrix having negative real parts, as long as the
quantity $x_i \equiv \beta_i p_i^2 B_i^* V_i^* = d_i ( b_i - e
\sum_{i=1}^m \frac{d_i}{\beta_i p_i} )$ is distinct for each $i$.

\begin{figure}[ht]
    \includegraphics[width = 1\columnwidth]{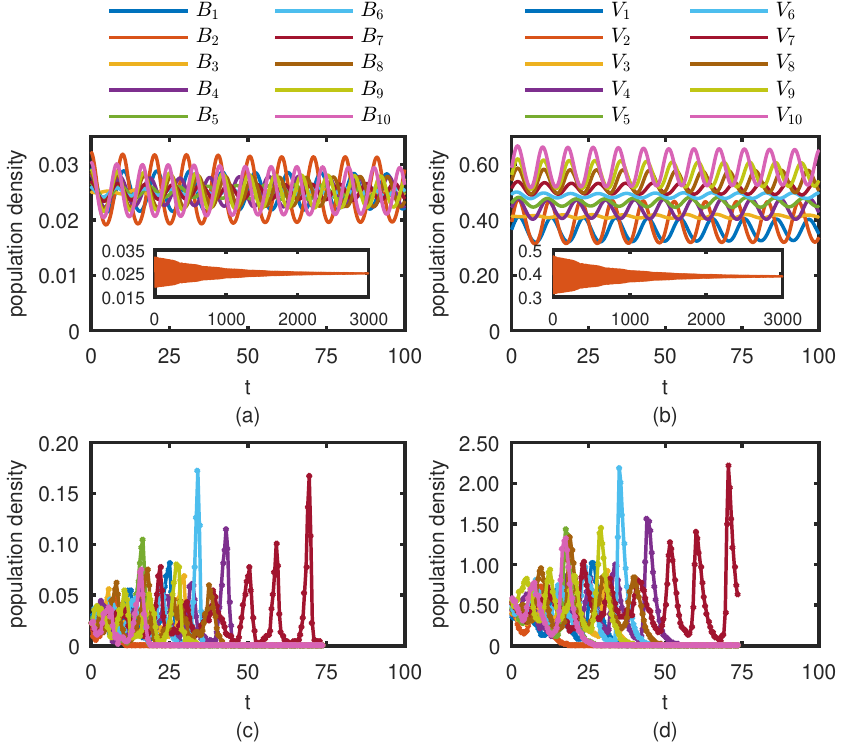}
    \caption{Population density time series obtained from the generalized KtW framework, with 10 bacterium-phage pairs.
    The left column is for bacteria and the right for viruses.
    The first row shows the result from a numerical evolution of the
    deterministic generalized KtW equations, with species densities
    initially perturbed randomly away from the steady state. The
    parameters are $\boldsymbol{b} = ($0.75, 0.8, 0.85, 0.9, 0.95, 1,
    1.05, 1.1, 1.15, 1.2$)$, $p_i \equiv p = 2$, $\beta_i \equiv \beta
    = 10$, $d_i \equiv d = 0.5$, and $e_{ij} \equiv e = 0.1$. Densities
    undergo oscillatory decay toward the steady state. The insets show
    the long time behavior which demonstrates that the steady state is
    a focus. For cosmetic reason, only the decays of $B_2$ and $V_2$
    are shown.
    The second row presents a typical simulation result of the
    stochastic version of the generalized KtW model, using the same set
    of parameters. The system size is $C=1000$ and populations are
    initialized with the steady state value. The oscillatory decay
    behavior is destroyed by demographic noise. Eventually, the system
    collapses after all bacterial strains become extinct
    around $t = 74$.} \label{multiPP_popu_t_fig}
\end{figure}

In Fig. \ref{multiPP_popu_t_fig}, we show in the first row the time
series of prey and predator densities obtained from a numerical
evolution of Eq. (\ref{multiPP_ODE}) for $m = 10$ pairs of bacteria and
phages. Species densities are initially perturbed away from the steady
state by a small random amount. As shown in the figure insets, species
densities decay back to the steady state at long times, confirming
the result of the linear stability analysis. The oscillatory behavior
on the short time scale is due to the imaginary parts of the
eigenvalues of the linear stability matrix.

To reveal the effect of demographic noise, we also conduct the
stochastic simulation of the corresponding individual level reactions
(\ref{multiPP_reactions}) with the same parameter set, using the
Gillespie algorithm \cite{gillespie1976}. The resultant species density
time series are shown in the second row of Fig.
\ref{multiPP_popu_t_fig}.
In contrast to the deterministic behavior of oscillatory decay, species
go extinct in a short time. Bacterial strains become extinct due to
random fluctuation; this consequentially triggers the extinction of the
corresponding viral strains, due to a lack of food. The number of
species monotonically decreases in the process, and the system
diversity undergoes a cascade.

The reason that the stable deterministic steady state of the
generalized KtW model cannot be maintained in the presence of
demographic stochasticity lies in the fact that species populations in
the stochastic model are all finite, and the probability of the
population reaching zero due to random drift is always nonzero.

Ecosystems have evolved many potential mechanisms to get around the
path to extinction, as introduced at the beginning of the article.
Here, we discuss one possibility: prey and predator coevolve with each
other so that fit mutants are constantly being introduced into the
system, thus preventing the elimination of the species. Specifically,
prey improve their phenotypic traits (e.g. strengthening the shell) to
escape from predators, and predators also adjust their corresponding
traits (e.g. sharpening the claws) to catch prey. This coevolutionary
arms race has been well-documented in many systems
\cite{bergelson2001plant_gene_for_gene,
chisholm2006plant_gene_for_gene,  person1959gene_for_gene,
thompson1992gene_for_gene, buckling2002Bac_virus,
buckling2002antagonistic_BV, raaberg2014MA_exp, luijckx2013MA_exp,
adema2015snail_specificity}. Previous theoretical studies focused on
the dynamics of the traits of prey and predator groups
\cite{Law1995evolutionary_cycling, Law1996dynamical_coevolution,
agrawal2002coevolution_model_spectrum, weitz2005coevolution}, and the
structure of the predation network
\cite{beckett2013coevolutionary_network}, under different coevolving
modes.
Here, we study how coevolution affects the diversity of the
host-specific predation system.

\smallskip
\noindent\textit{Coevolution model:-} We modify the stochastic
generalized KtW model (\ref{multiPP_reactions}) by adding in the
following two sets of reactions to describe mutations of the prey
$X_i$ from strain $i$ to $i\pm 1$, and similarly those of the predator
$Y_i$.
\begin{subequations} \label{coevolution_reactions}
\begin{align}
    X_i &\rightarrow X_{i\pm 1},  &\text{with rate }\mu_1/2, \\
    Y_i &\rightarrow Y_{i\pm 1},  &\text{with rate }\mu_2/2.
\end{align}
\end{subequations}
We assume that the mutation rates are strain independent and one
individual can mutate into its two neighbor strains with the same rate,
$\mu_1/2$ for bacteria or $\mu_2/2$ for viruses. We set the boundary
condition to be open, so that mutations out of the index set
$\{1,2,\dots, m\}$ are ignored. We will refer to Eq.
(\ref{multiPP_reactions}) and (\ref{coevolution_reactions}) as the
coevolving KtW (CKtW) model.

For sufficiently high mutation rates, the absorbing extinction state
in the generalized KtW model can be avoided, in the sense that a strain
can reemerge as mutants are generated from its neighbor relatives after its
population drops to zero. Therefore, mutation can stimulate a flux of
population through different strains and promote coexistence.

We define the diversity of the system in the CKtW model using the
Shannon entropy
\begin{equation}
    S = -\sum_{i=1}^m f_i \ln{f_i}.
\end{equation}
Here, $f_i $ is the fraction of the $i$th bacterial (viral) strain in
the entire bacterial (viral) community. The above expression reaches
the maximum, when all strains coexist at their deterministic steady
state values Eq. (\ref{multiPP_steady_state}), and the minimum $0$, when
only one strain exists. We score $S=-1$, if either the bacterial or
viral community goes extinct.

\begin{figure}[ht]
    \includegraphics[width = 1 \columnwidth]{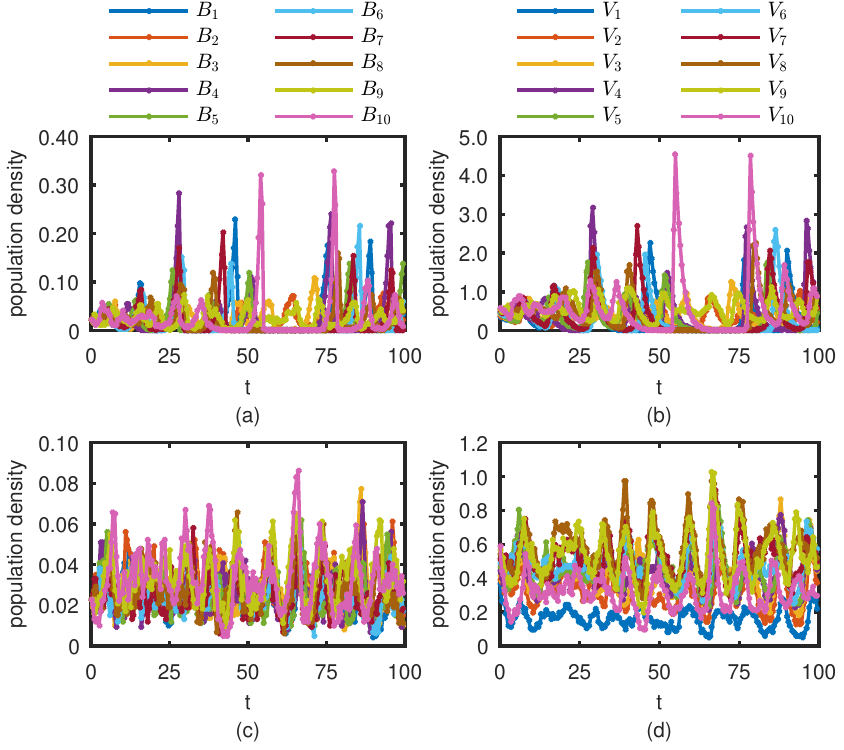}
    \caption{Population density time series in the stochastic coevolving KtW model.
    The left column is for bacteria and the right for viruses.
    The system size is $C=1000$, and the mutation rates are set to be equal, $\mu_1 = \mu_2 \equiv \mu$. All other rates are the same as those in Fig. \ref{multiPP_popu_t_fig}.
    The first row shows the case of a small mutation rate $\mu = 0.015$. Populations undergo a temporal winner alternation.
    The second row is obtained with a high mutation rate $\mu = 1$. Strains coexist, with small fluctuations around the steady state.} \label{KtW_coevolution_popu_t}
\end{figure}

\begin{figure}[ht]
    \subfloat[]{%
        \includegraphics[width = 1 \columnwidth]{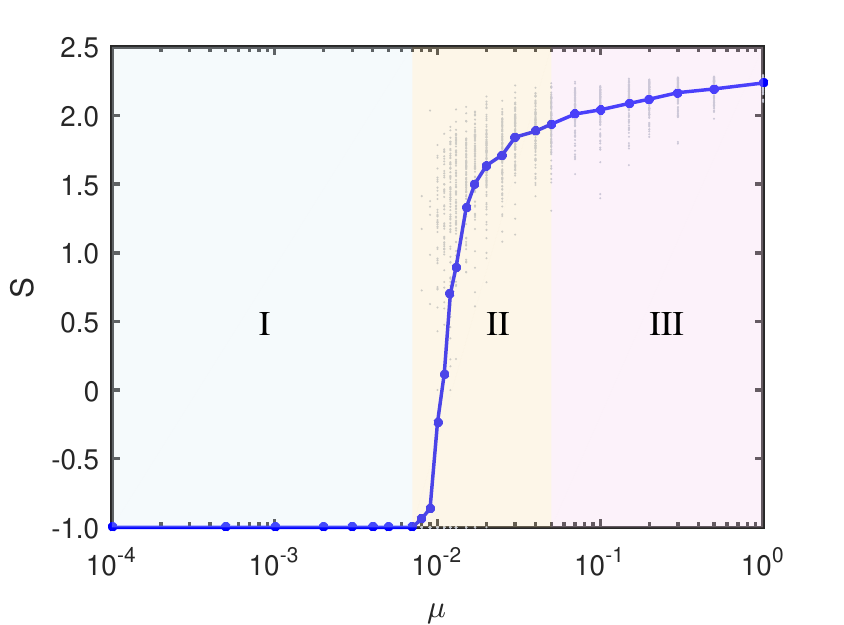}}\\
    \subfloat[]{%
        \includegraphics[width = 1\columnwidth]{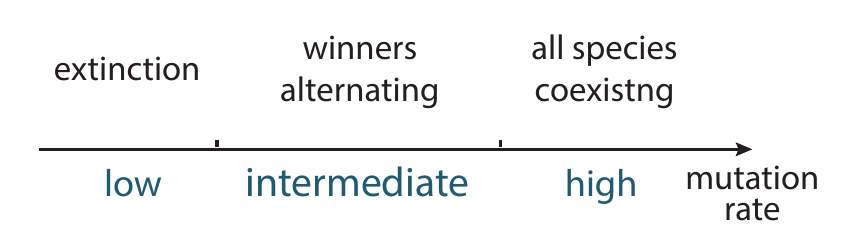}}
    \caption{(a) The prey diversity, represented by the Shannon entropy $S$, as a function of the mutation rate $\mu_1 = \mu_2 \equiv \mu$. For each value of $\mu$, we conduct $100$ replicates of simulations. The diversity in the replicate is assigned to be $-1$, if either the prey or predator community goes extinct by the end of the simulation. Otherwise, the Shannon entropy is calculated at the end of the simulation. The gray dots in the figure represent the diversity values in all replicates. The blue dots are the average values of diversity at each $\mu$.
    For this particular set of parameters, the mean-field generalized KtW equations give equal bacterial strain concentration at the steady state, and the maximum diversity in the corresponding CKtW model is $\ln m$.
    (b) A descriptive phase diagram of the dynamics, with the mutation rate as the tuning parameter.
    } \label{KtW_coevolve_Shannon_u_fig}
\end{figure}

We present population density time series in Fig.
\ref{KtW_coevolution_popu_t}, and the dependence of prey diversity on
the mutation rates in Fig. \ref{KtW_coevolve_Shannon_u_fig}. We set
$\mu_1 = \mu_2 \equiv \mu$ for simplification. For small enough
mutation rate (time series not shown), the entire community can go
extinct before mutants can emerge, and the system still collapses as in
the generalized KtW model. This corresponds to region I in Fig.
\ref{KtW_coevolve_Shannon_u_fig}(a). For intermediate mutation rates,
most strains stay near extinction, driven by demographic noise, while
some mutants can grow to a dominant status if they happen to confront
only a few predators when first emerging. Subsequently, the predator
population expands, feeding on the dominating winners, thus reducing
the winner population, and allowing the next dominator to grow. In this
way, we see that winner populations spike alternatively in the time
series, as in the first row of Fig. \ref{KtW_coevolution_popu_t}. Near
the onset of coexistence, the diversity has a large deviation and is
very sensitive to the mutation rate, as shown in region II in Fig.
\ref{KtW_coevolve_Shannon_u_fig}(a). For large mutation rate, the
coevolution-driven population flow is fast enough to compensate for the
demographic fluctuations. All strains remain near the steady state, and
no one can win over others, as shown by the population time series in
the second row of Fig. \ref{KtW_coevolution_popu_t}. The diversity
approaches slowly the maximum, with small deviations, as demonstrated
in region III in Fig. \ref{KtW_coevolve_Shannon_u_fig}(a). For
extremely large mutation rate (time series not shown), we can not view
the mutation as a perturbation to the ecological population dynamics.
Species populations will deviate from the mean-field steady state Eq.
(\ref{multiPP_steady_state}) due to the large effect of mutations.
According to the above discussion, there are three phases of dynamics,
as illustrated in Fig. \ref{KtW_coevolve_Shannon_u_fig}(b), the
extinction phase at low mutation rate, the winner-alternating phase at
intermediate mutation rate, and the coexisting phase at high mutation rate.

\smallskip
\noindent\textit{Open system model:-} Up to now, the models are closed:
we have pre-assigned the number of predator-prey pairs in the system,
and furthermore set boundaries to the mutation of species. A more
realistic approach is to let the system be open, and evolve by itself
to establish however many species there can be.

Suppose that a fit species mutates, by changing its
phenotypic traits to escape from its predator. As mutants take on new
traits, the population spreads in trait space. This expansion usually
is associated with a trade-off in the fitness
\cite{Law1995evolutionary_cycling}: the further the trait is from the
origin, the lower the growth rate becomes. We model the trade-off
effect by assigning species discretely at  certain trait values, so
that the species index basically represents the trait. Specifically, we
assume the trait space is 1-dimensional and set up $M$ species in it;
we assign the highest birth rate to the species with index $M/2$, which
is at the center of the trait space and then is the origin of the trait
expansion; we decrease the birth rate as the species index goes from
$M/2$ to $1$ and and from $M/2$ to $M$, manifesting the trade-off of
mutation. Species $1$ and $M$ have the lowest birth rates that are
almost $0$, and further mutation of the two will result in mutants with
negative birth rates, which can not grow and are thus excluded from the
model. This species space $\{1,2, \dots, M\}$ contains
all possible species that can potentially exist in the system. However,
under conditions of resource limitation, only a few with relatively high
growth rates, out of $M$, can eventually be established in the system.
In our model, this limit due to the carrying capacity is set by the
competition strength $e$. The number of species that manage to thrive
corresponds to $m$ in the previous models.

We conduct simulations to test the dependence of diversity on the
mutation rate. See the Supplemental Material for the resultant
population time series and diversity dependence on the mutation rate.
Even though the number of established pairs varies with time and the
population leaks out of the region deterministically allowed by the
carrying capacity, the system still exhibits three phases depending
on the mutation rate, similar to the CKtW model with fixed number of
species.

\smallskip
\noindent\textit{Discussion:-}
In the intermediate and fast mutation regions of the CKtW model, the
ecological and evolutionary dynamics are coupled to each other and
occur on the same time scale \cite{yoshida2003rapid}. This type of
coupling can most easily be observed in microbial systems, in which
organisms have a high mutation frequency
\cite{bohannan1997effect,shih2014rapid_evolution,yoshida2007cryptic}.
Recent work has shown clearly the existence of genomic islands, where
genomes of different strains vary in loci thought to be associated with
phage resistance \cite{Rohwer2009genomic_island}. Both host-specific
predation and mutation are important in generating the observed
diversity of the bacterial genome. The minimal CKtW model can in
principle describe the diversity in the above system. For example, by
controlling the mutation rate through an inducible promoter, using
molecular techniques pioneered in Ref. \cite{kim2016real}, we envisage
a fast bacterium-phage coevolution experiment to test the phase diagram
predicted by our model.

In addition to inevitable simplification of biological details, both
the generalized KtW and the coevolving KtW models assume that the
system is well mixed, ignoring any spatial dispersion. Consequently,
they can not capture the reservoir effect \cite{shmida1984coexistence}
present in an ecosystem, which means that for any local community,
organisms in its surrounding environment can move into it, keeping it
supplied and refreshed. Specifically, even if a species goes extinct in
a local community, it can be reseeded there by the surrounding
reservoir. Well-mixed models should be thought of as describing not the
entire system, but a much smaller correlation volume, in which local
demographic stochasticity can be significant
\cite{butler2011fluctuation,tauber2012population,biancalani2017non_normality}.

\medskip
\begin{acknowledgments}
\noindent\textit{Acknowledgments:-} This work was partially supported
by the National Science Foundation through grant PHY-1430124 to the NSF
Center for the Physics of Living Cells, and by the National Aeronautics
and Space Administration Astrobiology Institute (NAI) under Cooperative
Agreement No. NNA13AA91A issued through the Science Mission
Directorate.
\end{acknowledgments}

\bibliography{v1.6_KtW_coevolution_ref}
\bibliographystyle{apsrev4-1}

\end{document}